\documentstyle[11pt,psfig]{article}

\def\FF{\hbox to 8.33887pt{\rm I\hskip-1.8pt F}}
\def\NN{\hbox to 9.3111pt{\rm I\hskip-1.8pt N}}
\def\PP{\hbox to 8.61664pt{\rm I\hskip-1.8pt P}}
\def\QQ{\rlap {\raise 0.4ex \hbox{$\scriptscriptstyle |$}}
{\hskip -4.5pt Q}}
\def\RR{\hbox to 9.1722pt{\rm I\hskip-1.8pt R}}
\def\ZZ{\hbox to 8.2222pt{\rm Z\hskip-4pt \rm Z}} 
\font\tenfrak=eufm10\font\sevenfrak=eufm7\font\fivefrak=eufb5
\newfam\frakfam
     \textfont\frakfam=\tenfrak
     \scriptfont\frakfam=\sevenfrak   
     \scriptscriptfont\frakfam=\fivefrak
\def\frak{\fam\frakfam\tenfrak}

\def\cartprod{\mathop{\lower2pt\hbox{{\twelvesans X}}}}

\def\bbar#1{\setbox0=\hbox{$#1$}%
     \copy0\kern-\wd0
     \raise.0433em\box0}
\def\optbar#1{\vbox{\ialign{##\crcr\hfil${\scriptscriptstyle(}\mkern -1mu
         \vrule height 1.2pt width 3pt depth -.8pt
         {\scriptscriptstyle)}$\hfil\crcr
          \noalign{\kern-1pt\nointerlineskip}$\hfil\displaystyle{#1}\hfil$\crcr}}}
\def\<{\left<}
\def\>{\right>}

\def\sbar#1{\vbox{\ialign{##\crcr
         \noalign{\kern1pt\nointerlineskip}
         \hfil$ \mkern -1mu\vrule height 1.2pt width 3pt depth -.9pt$\hfil\crcr
         \noalign{\kern0pt\nointerlineskip}
                $\hfil{\sst #1}\hfil$\crcr}}}

\newcommand{\bequ}{\begin{equation}}

\newcommand{\eequ}{\end{equation}}

\newcommand{\barr}{\begin{array}}

\newcommand{\earr}{\end{array}}

\newcommand{\R}{{\rm l\kern -1.6pt{\rm R }}}

\def\al{\alpha}

\def\de{\delta}
\def\ep{\epsilon}

\def\et{\eta}

\def\ka{\kappa}
\def\la{\lambda}

\def\si{\sigma}

\def\ph{\phi}
\def\ch{\chi}
\def\ps{\psi}

\def\La{\Lambda}
\def\Si{\Sigma}

\def\bbbone{{\mathchoice {\rm 1\mskip-4mu l} {\rm 1\mskip-4mu l}    
{\rm 1\mskip-4.5mu l} {\rm 1\mskip-5mu l}}}
\long\def\prf{\removelastskip\penalty-100\vskip\baselineskip\noindent{\bf
            Proof:\ }\ \ \ignorespaces}
\def\sq{\hbox{\rlap{$\sqcap$}$\sqcup$}}
\def\qed{\ifmmode\sq\else{\unskip\nobreak\hfil
           \penalty50\hskip1em\null\nobreak\hfil\sq
           \parfillskip=0pt\finalhyphendemerits=0\endgraf}\fi}

\def\cS{{\cal S}}
\def\Pn{{{\cal P}_n}}

\def\gF{{\frak F}}
\def\gT{{\frak T}}

\begin{document}
\title{Constructive Renormalization Theory}
\author{Vincent Rivasseau\\
Centre de Physique Th{\'e}orique, CNRS UPR 14\\ 
Ecole Polytechnique, 99128 Palaiseau Cedex, FRANCE\\
email: rivass@cpht.polytechnique.fr}

\date{}
\maketitle

\thispagestyle{empty}
\begin{abstract}

These notes are the second part of a common course on Renormalization 
Theory given with Professor P. da Veiga\footnote{X Jorge Andr{\'e}
Swieca Summer School, Aguas de Lind{\'o}ia, February 7-12, Brazil.}. 
I emphasize here the rigorous non-perturbative or constructive aspects of 
the theory. The usual formalism for the renormalization group in field 
theory or statistical mechanics is reviewed, together with its limits. 
The constructive formalism is introduced step by step. Taylor
forest formulas allow to perform easily the cluster
and Mayer expansions which are needed for a single step of
the renormalization group in the case of Bosonic theories.
The iteration of this single step leads to further
difficulties whose solution is briefly sketched. The second part of the 
course is devoted to Fermionic models. These models are easier to treat
on the constructive level, so they are very well suited to beginners
in constructive theory. It is shown how the Taylor forest formulas allow
to reorganize perturbation theory nicely in order to construct the
Gross-Neveu$_{2}$ model without any need for cluster or Mayer expansions. 
Finally applications of this technique to condensed matter and renormalization
group around Fermi surface are briefly reviewed.

\end{abstract}

\section{The Renormalization Group: an overview}

\subsection{Functional Integration and its problems}
In this section we restrict ourselves to the bosonic
$\phi^4$ field theory in $d$ Euclidean space time dimensions.
The model, introduced in P. Da Veiga's lectures, is defined by the
(formal) measure
$$ d\mu_{C} (\phi ) e^{-S(\phi)}  , \quad S(\phi) = \la \int \phi^4 (x) d^d x
\eqno({\rm 1.1})
$$
where $d\mu_{C} (\phi )$ represents the Gaussian measure for the free
field. Gaussian measures are characterized by their covariance,
or propagator, which for a massive theory is, in Fourier space:
$$ C (p)  = (p^2 + m^2 )^{-1} \ ,  \eqno({\rm 1.2})
$$
and $S$ is the (bare) action. In dimension $d=2,3$ the model is 
superrenormalizable, and its rigorous construction was the first 
major achievement of constructive theory [GJ].
In dimension $d=4$ the model is just renormalizable,
hence a more general action, including mass and wave
function counterterms is needed:
$$ S(\phi) = \la \int \phi^4 (x) d^d x + \mu \int \phi^2 (x) d^d x
+ a \int (\nabla\phi) ^2 (x) d^d x
\eqno({\rm 1.3})
$$
We recall the three main problems associated with giving a mathematical
precise sense to the formal measure (1.1).

\medskip
\noindent {\bf A) The ultra-violet problem} 
The propagator (1.2) has not a well defined
kernel in direct space for $d\ge 2$, since $ \int d^d p
(p^2 + m^2 )^{-1}  $ is not absolutely convergent. This does not prevent
mathematically to define the Gaussian measure associated to this propagator,
for instance through Minlos theorem [GJ], since the propagator is positive
in Fourier space.
However it entails that this measure is supported by {\it distributions}.
The immediate consequence is that a local interaction like $\phi^4 (x)$
is ill defined, since multiplication of distributions in general
is ill-defined. Therefore we must truncate the propagator at high
momenta by means of a cutoff, introducing for instance
$$  C_{\kappa} (p) = \int _{\kappa}^{\infty} e^{- \alpha  (p^2 + m^2 )} 
d \alpha \eqno({\rm 1.4})
$$
which has a well defined kernel 
$$  C_{\kappa} (x, y) = \int d^{d}p e^{ip.(x-y)}
\int _{\kappa}^{\infty} e^{- \alpha  (p^2 + m^2 )} 
d \alpha   = \int _{\kappa}^{\infty} e^{- \alpha m^2 -  |x-y|^2 /\alpha } 
{d \alpha \over \alpha^{d/2}} \eqno({\rm 1.5})
$$
and to let $\kappa \to 0$ later.

As a consequence the support of $d\mu_{C_{\kappa}} (\phi )$ is now made of
perfectly smooth functions, and  $\phi^4 (x)$ becomes well-defined.
An other popular ultraviolet cutoff, especially for gauge theories,
is the lattice cutoff, but I will not use it here.

\medskip
\noindent {\bf B) The infra-red problem} 
Even with an ultraviolet cutoff, if $\phi^4 (x)$ is well defined,
with probability one it is not decreasing at infinity, hence it is
certainly not {\it integrable} on $\R^{d}$. Hence the quantity
$S(\phi) = \int_{\R^{d}} \phi^4 (x) d^d x $ with probability 1 is 
ill-defined. The solution is to restrict the interaction to a finite 
volume (usually a box $\Lambda$ of size $L$), and to let $L \to \infty$
later.

The measure with these two cutoffs becomes
$$ d\mu_{C_{\kappa}} (\phi ) e^{- \la \int_{\Lambda} \phi^4 (x) d^d x}
\eqno({\rm 1.6})
$$

\medskip
\noindent {\bf C) The large field problem}
In (1.6) the integrand is now well defined on a set of measure one. 
But it does not mean that it can always be integrated!
Even in one-dimensional integration, $F(\la) = \int_{-\infty}^{+\infty}
e^{-x^2/2 - \la x^4 } dx $ converges only for $\la \ge 0$.
We cannot hope the infinite dimensional integral (1.6) to be better
behaved than this one dimensional integral. Therefore Bosonic 
functional integrals require some stability for the potential
(here e.g. $\la \ge 0$). As dicussed by P. da Veiga, the perturbation
series for $F$ do not converge. $F$ is not analytic, but Borel
summable. This is also the best we can hope for the $\ph^4$ theory,
and what has been proved in dimension 2 or 3.

Convergence of the functional integral itself, and the divergence
of perturbation theory can be considered as generic ``large field''
problems, because they are related to the fact that a bosonic field
is an unbounded variable. Physically a large field corresponds
to a large number of excitations or particles being produced, and
large field problems are generic in Bosonic theories because Bosons, 
instead of Fermions, can pile up in large
numbers at the same place.

\subsection{Thermodynamic limit}
Having reviewed the main mathematical problems of field theory
in the functional framework, we remark
that cutoffs by themselves alone cannot solve any 
problem. For instance the theory with cutoffs does not satisfy any
reasonable axioms. Some manipulations have to be performed,
{\it cancellations} have to occur, so that the limits $\kappa \to 0$
and $L\to \infty$ which looked hopeless at first sight 
become finite and well defined.

The conceptually simplest of these manipulations is the thermodynamic
limit $\Lambda\to \RR^{d}$, which is particularly easy
when the theory has a mass. We know from statistical mechanics
that only the ``intensive'' or thermodynamic quantities should have a limit 
as $\Lambda\to \infty$, the extensive ones, proportional to the volume,
should diverge. Hence the only ``manipulation'' in this case is to
restrict our attention to quantities such as the pressure $p$
or the normalized Schwinger functions $S_{n}$ of the theory:

$$ p = \lim_{\Lambda \to \infty}   {1\over |\Lambda|} \log Z(\Lambda);
\quad   Z(\Lambda) =  \int   e^{- S_{\Lambda} (\phi)} d\mu_{C} (\phi )
\eqno({\rm 1.7})$$
$$ S_{N} = \lim_{\Lambda \to \infty} {1\over Z(\Lambda) } \int \phi (x_{1})
... \phi (x_{N})  e^{- S_{\Lambda} (\phi)} d\mu_{C} (\phi )
\eqno({\rm 1.8})$$
In these quotients both quantities should diverge as $\Lambda\to \infty$,
but their ratio should have a finite limit. This is clear in perturbation
theory: the series for extensive or unnormalized quantities correspond
to general graphs, which may contain in particular vacuum 
graphs which by translation invariance,
give rise to infinite integrals when the spatial integration
is over all $\RR^{d}$. The intensive quantities such as
$p$ or $S_{N}$ have power series made of connected graphs, with at least
one vertex fixed, so that if the propagator itself is integrable, as 
is the case in a massive theory, the corresponding amplitudes are finite
in the thermodynamic limit. In the next sections we will see how the
thermodynamic limit can be performed non-perturbatively.
For Bosons it requires a cluster and a Mayer expansion, whereas for Fermions,
it requires almost nothing: 
simply reorganizing the finite dimensional integrals of
perturbation series in terms of trees rather than Feynman graphs.

\subsection{Renormalization}

The ultraviolet problem increases in difficulty
when $d$ increases. In general we cannot
expect a finite $\kappa \to 0$ limit for quantities such as the Schwinger
functions if we keep the bare parameters fixed as $\kappa \to 0$. 
But physically these bare parameters are not observable in low energy
realistic accelerators! P. da Veiga has already explained, at least at the
perturbative level, how to get rid of the ultraviolet divergences
that occur in the Feynman graphs of the theory. For $d=4$
where the theory is just renormalizable, perturbation theory suggests
that if we allow the {\it bare} parameters
of the theory $\lambda$, $\mu$ and $a$ in (1.3) to depend on the cutoff 
$\kappa$ in such a way as to diverge
as $\kappa \to 0$, cancellations can take place so that the Schwinger
functions  $S_{N}$ might have a well defined
limit as $\kappa \to 0$. I say that it only suggests this, since
bare quantities in standard perturbative renormalization are only
defined as formal power series in the renormalized coupling.
These series are not necessarily summable, as explained in da Veiga's lectures,
both because of the large number of graphs, and because perturbative 
renormalization creates additional sources for divergences, namely
renormalons. It is very risky to speculate whether or 
not an infinite non summable series of diverging terms 
really produces a quantity
diverging with the cutoffs (in fact in asymptotically
free cases it does not!).

I will not add further remarks to the lectures
of da Veiga on perturbative renormalization, except to
recall that historically many mathematicians felt
renormalized perturbation 
theory ``pulls the infinities of quantum field theory under the rug''
by hiding them into the bare constants.
Physicists also felt in the early days of renormalization
theory that it was not ``natural''; in particular it did not explain
why nature seems to prefer renormalizable field theories. Professor Dyson,
one of the founding fathers of renormalization, once told me that for him the 
main surprise of the theory was that it lived so long!
In short perturbative renormalization, although beautiful and encouraging,
and certainly useful for many phenomenological computations (QED, QCD in some
regimes...) is a partial and unsatisfactory answer to the problems of 
quantum field theory both for mathematicians and for physicists.

\subsection{The Renormalization Group}

A key progress in the theory of renormalization was the introduction
of the philosophy of the Renormalization Group (hereafter called RG)
by Wilson and followers.
The basic idea is the following: since the
limit $\kappa \to 0$ is so hard to grasp,
let us not perform it at once, but in steps. A single step can be well
understood, and the problem is reduced to the hopefully simpler
problem of iterating many times a relatively simple transformation.
Hence RG does not solve renormalization, 
but it replaces it by a problem in dynamical systems. This seems an almost
trivial idea, but it is really also a tremendous change in perspective which 
immediately led to progress. For instance it replaced the old ``Landau 
ghost'' or its modern version, the renormalon problem,
by a different question: is the theory asymptotically free (or safe), 
so that the flow for the coupling constant remains bounded at all scales?
Also the renormalization group philosophy can be
applied directly to {\it infrared} problems and statistical
mechanics with a lot of success.

As you surely know, the $\phi^{4}_{4}$ theory for the right sign of the
coupling constant is not asymptotically free.
At least until now this means that we do not know
how to perform its limit $\kappa \to 0$ and end up with
a non trivial interacting theory (although 
$\phi^{4}_{4}$ remains interesting for
pedagogical discussions). But the good news are that other theories such 
as gauge theories or the two dimensional Gross-Neveu and
$\si$ models are asymptotically free. Also the renormalization group explained
quickly why nature seemed to prefer renormalizable theories: indeed
in a generic interaction at a very high momentum scale, 
non renormalizable terms are irrelevant; washed out by the RG flow, their 
presence cannot be detected in the effective theory at low energy. 

Let us briefly sketch what would be
the RG strategy for the $\phi^{4}_{4}$ theory:

The key idea is to split the space of all frequencies into
discrete slices, following a geometric
progression. This can be done on lattices with
the popular tool called block-spinning, but we can also simply
cut the propagator into ``momentum slices''. Taking some fixed ratio $M$
for the geometric progression, and $\kappa = M^{-2\rho}$, $\rho$
being an integer, we have:

$$  C_{\rho} (p) =  \sum_{j=0}^{\rho} C^{j}(p) 
\eqno({\rm 1.9})$$
$$C^{0}(p) =
\int _{1}^{\infty} e^{- \alpha  (p^2 + m^2 )} 
d \alpha \quad ; \quad C^{j}(p) =
\int _{M^{-2j}}^{M^{-2(j-1)}} e^{- \alpha  (p^2 + m^2 )} 
d \alpha  \ {\rm for } \ j \ge 1
\eqno({\rm 1.10})$$
There is a corresponding separation of the field into a sum of
independent random variables: $\phi = \sum_{j} \phi_{j}$,
$\phi_{j}$ distributed according to $C^{j}$. $\phi_{j}$ is called
the ($j$-th) fluctuation field and 
$\sum_{k=0}^{j-1} \phi_{k}$  the ($j$-th) background field.

\medskip\noindent{\bf Exercise 1.1} 
Prove that for some constant $K$:
$$ C^{j}(x, y) \le K M^{2j} e^{- M^{j} |x-y|/K}
\eqno({\rm 1.11})$$
$\hfill \clubsuit$\vspace{.15cm}

Now introduce the operation $*$ by
$$ \mu_{j} * Z (\phi) = \int d\mu_{C^{j}} (\zeta) Z (\phi + \zeta)
\eqno({\rm 1.12})$$
where $\zeta$ plays the role of a fluctuation and $\phi$ of a 
background field.

If we define $S_{\rho}(\phi)$ as the bare action and $Z_{\rho}(\phi)= 
e^{-S_{\rho}(\phi)}$, we have:

$$ Z = \int d\mu_{C_{\rho}}(\phi) Z_{\rho}(\phi) = 
(\mu_{0} *...*(\mu_{\rho -1}*
(\mu_{\rho}* Z_{\rho})
\eqno({\rm 1.13})$$ 

Let us define 
$$ Z_{j} = (\mu_{j} *...*(\mu_{\rho - 1}*
(\mu_{\rho}* Z_{\rho})\quad ;\quad  Z_{j-1} = \mu_{j-1} * Z_{j}; 
\eqno({\rm 1.14})$$
and 
$$ S_{j} (\phi) = -\log Z_{j} (\phi)  \ .
$$
We see that constructing the ultraviolet limit is the same as
finding a bare action $S_{\rho}(\phi)$ such that the effective,
or renormalized action $S_{0}(\phi)$ converges as $\rho \to \infty$.
 
Remark that the $C^{j}$ satisfy approximate scaling laws:
$C^{j+1} (x) \simeq  M^{2}   C^{j} (Mx) $.
They would even satisfy exact scaling if we consider
a massless theory with $m=0$ which is often done. In this 
case we can use the scaling $C^{j+1} (x) = C^{j} (Mx)$, to
perform the famous trivial but slightly confusing rescalings
of the renormalization group. Defining 
$\phi_{M} (x) =  M^{(d-2)/2} \phi (Mx)$, and
$ \widetilde Z_{j} (\phi) =  Z_{j} (\phi_{M^{j}}) $
we obtain indeed the equation
$$ \widetilde Z_{j} (\phi)= (\mu_{1} * \widetilde Z_{j+1})(\phi_{M^{-1}})
$$
This defines the ($j$ independent) ${\cal R}$ operation on the 
action $S$ as the composition of 4 steps:

-spatial rescaling by $M^{-1}$

-field rescaling by $ M^{-(d-2)/2}$

-integration over a fixed (scale 1) fluctuation slice

-taking the logarithm of the result to reexpress it as an action for 
the next step. 

In this way the problem of the ultra-violet limit reduces to 
convergence of the $\rho$ times 
iterated transformation ${\cal R} o {\cal R} ... o {\cal R} S_{\rho}$ 
as $\rho \to \infty$.
The problem of iteration of a fixed map can be analyzed as a
discrete dynamical system: if a fixed point appears, for instance
the free field or a theory close to it, one can hopefully 
trust a perturbative analysis of the vicinity of this fixed point.

In particular we see that for instance polynomial terms such as
$\int \ph^{n}(x) d^4 x$ scale with a factor (corresponding to
power counting) which is $M^{(4-n)j}$ after $j$ steps. For $n>4$
they are ``irrelevant'' and die out in the RG flow; the $\ph^4 $
term is marginal, and its flow is governed by the sign of the ``bubble
graph'', hence by the sign of the beta function at small coupling.
Finally the mass term, quadratically growing, is ``relevant''.
Adding derivative couplings, we find that only the wave function
term $\int (\nabla\ph)^{2}(x) d^4 x$ is not irrelevant, but marginal.
From this very simple analysis follows the classification of renormalizable
theories into asymptotically free (Gaussian fixed points)
and not asymptotically free,
and the discovery of non-Gaussian fixed points of the RG close to Gaussians,
if one modifies for instance slightly the canonical scaling of 
the fields (mimicking non-integer dimensions) for an asymptotically free model.

Study of infrared critical points by RG is very similar to that of ultraviolet
limits, except that in an infrared problem,
the ultraviolet cutoff is fixed, hence it is natural
to give it an index 0 or 1; and the slices
run towards momentum
$p=0$, so that as $j$ grows, it indexes smaller and smaller momenta.

Sometimes here a confusion arises: the renormalization group
does not simply ``exchange'' infrared and ultraviolet limits.
The basic fact to keep in mind is that it flows
always in the same direction: from small spatial
scales to larger ones. Indeed it deduces macroscopic actions
from microscopic ones, as is traditional in the physical analysis
of a phenomenon, where large scale effects are explained by smaller ones
\footnote{We are aware that this traditional view is put in question 
by duality in string theory, but in this lecture we nevertheless stick to it!}.
Hence the direction of flow of RG never changes, only the
point of view of the observer. In an
ultraviolet problem in some sense it is the source of the RG river
which goes to infinity, whereas in an infrared problem it is the
mouth of the river which goes to infinity relatively to the observer. 
 
\subsection{Constructive RG is necessary!}

After this blitz overview of the standard renormalization group of textbooks 
remark that although it clarifies enormously the 
ultraviolet problem, it is not yet formulated in a correct mathematical
way. In particular starting form any polynomial action it 
creates an effective action which is obviously no longer polynomial,
and this even after a single step! 
Therefore the large field problem (integration on $\phi$, or $\zeta$), 
appears! More precisely the behavior of $S_{eff} (\phi)$
at large $\phi$ is unclear, so that starting
from a stable interaction, even the second step of the RG may be already
ill-defined. This point has to be stressed to  physicists!

The solution is to reconsider the single step of the RG, hence the
theory in a finite slice with a background field and to analyze it
in a way which can be iterated correctly mathematically. This is the
``constructive" version of the RG, as introduced in the 80's by
Gallavotti and coworkers {BCGNOPS}, and developed in [GK1-2].
It has been now well developed and systematized
into a coherent mathematical formalism, by Pordt, 
Brydges and collaborators [P][B]. A related formalism is 
the ``multiscale phase space expansion'' of [FMRS1-2], now redesigned 
into a more transparent and explicit formalism in [A][AR2].
The main difference is that in the phase space expansion formalism
the rescalings are not performed, so that the iteration of
the RG steps is fully developed. Phase space expansion therefore
leads to an explicit description of the solution of the RG induction, somewhat
like Zimmermann's forest formulas ``solves'' Bogoliubov's induction. 

In these formalisms the main tool consists in performing
the key step of integration of fluctuation (the mathematically
well defined version being called a ``cluster expansion'') and 
taking the logarithm (the mathematically
well defined version being called a ``Mayer expansion'') only
in regions where the background field is small. In the other so-called
large field regions one simply ``waits and sees''.
Although a full presentation of the constructive RG formalism for bosonic 
theories is beyond the scope of this course, we give 
in the next section in detail
the treatment of a single scale model with a cluster + Mayer expansion.

The constructive RG will be an iteration of this treatment, with
the added complication (alas very nontrivial) that one cannot expand 
the ``large field regions'' and must exploit the fact that  
their weights are small in the probabilistic sense.

Using the constructive RG approach or the related ``multiscale" expansions
some models have been built, giving the first
concrete examples of renormalizable quantum field theories
fulfilling all Wightman axioms, such as the Gross-Neveu$_{2}$ model [GK1]
[FMRS1].
New models not perturbatively renormalizable
but asymptotically safe become also accessible, such as the Gross-Neveu$_{3}$ 
model built by P. da Veiga and collaborators [dV].
In the infrared regime
bosonic models of renormalizable power counting such as the critical
(massless) $\phi^{4}_{4}$ with an infrared cutoff [GK2]
[FMRS2], or 4 dimensional weakly
self avoiding polymers have been controlled [IM1], 
and their asymptotics at large
distance established. Nonperturbative mass generation has been established
in the Gross-Neveu$_{2}$ model and in the nonlinear 
$\sigma$ model at large number of components [KMR][K].
Finally the RG when applied to condensed matter give rise to many 
rigorous results and programs, and this is the subject of the last section.

\section{Single Scale Bosonic Model: the cluster and Mayer expansions}

\subsection{Tree and Forest formulas} 

In statistical mechanics, the key step of the thermodynamic limit
is to take a logarithm to pass to intensive quantities which correspond
to connected quantities. The minimal discrete connecting structures
between points are trees. It is therefore worth devoting some time to
the precise combinatorics of trees, and to the way to generate them 
systematically between points.

Let $n\ge 2$  be an integer, $I_n=\{1,\ldots,n\}$, 
$\Pn=\bigl\{\{i,j\}/i,j\in I_n,i\ne j\bigr\}$ (the set of unordered pairs
in $I_n$).
An element $l$ of $\Pn$ will be called a {\sl link}, a subset of $\Pn
$, a {\sl graph}. A graph $\gF=\{l_1,\ldots,l_{\tau}\}$ containing no loops,
i.e.{} no subset $\bigl\{\{i_1,i_2\},\{i_2,i_3\},\ldots,\{i_k,i_1\}\bigr\}$
with $k\ge 2$ elements, is called a forest. A connected forest is called
a tree.
A forest is a union of disconnected trees $\gT$, the supports of which are 
disjoint subsets of $I_{n}$ called the 
{{\sl connected components} or {\sl clusters} of $\gF$. 

The first basic result on trees goes back to the XIXth century:

\medskip
\noindent{\bf Theorem 1 (Cayley's Theorem)} {\it 
There are exactly $n^{n-2}$ labeled trees on
a set of $n$ points.}
\medskip

\noindent {\bf Exercise 2.1:} Prove Cayley's theorem by induction,
using the multinomial expansion, and showing that there are
$n!/\prod(d_{i}-1)!$  trees with coordination number
$d_{i}$ at vertex $i$, using the fact that trees always have ``leaves''.
 $\hfill \clubsuit$\vspace{.15cm}

Let $\cS$ be the space of smooth functions of
$n(n-1)/2$ variables ${\bf x}={(x_l)}_
{l\in\Pn}$ associated to the links of $\Pn$.
A Forest formula is a Taylor formula with
integral remainder, which expands a function $H$ of $\cS$
to search for the explicit
dependence on given link variables $x_{l}$. Taylor formulas with remainders
in general are provided with a ``stopping rule'' 
and forest formulas stop at the level of connected sets. This means that
two points which are joined by a link are treated as a single block.
(More sophisticated formulas with higher stopping rules
are very useful for higher particle irreducibility analysis, 
or renormalization group computations
but are no longer forest formulas in the strict sense [AR2]). 

Any forest is a union of connected trees. Therefore
any forest formula has an associated tree formula for its
connected components. And therefore, at least formally,
any forest formula solves the problem of computing normalized 
correlation functions. Indeed applying the forest formula to the 
functional integral for the unnormalized functions, the connected functions
are simply given by the connected pieces of the forest
formula, hence by the corresponding tree formula. It is in this sense
that forest formulas exactly solve the well known snag that 
makes connected functions difficult to compute. This snag is that since
typically there are many trees in a graph, one does not know ``which one to
choose'' when one tries to compute connected functions in the (desirable) 
form of tree sums. Any forest formula gives a particular answer to that 
problem, associating a weakening factor $w$ to the links which remain 
underived (the potential ``loops''). This weakening factor $w$  tells us 
exactly by how much our pondered ``tree choice'' has ``weakened'' 
the remaining loop lines.

Several forest formulas exist, with different 
rules for $w$; they correspond 
to different ways of letting forests grow.
In one logic, the most ``symmetric'' one,
the forest grows in a random way: it leads to a formula
first established by Brydges and Kennedy [BK]. In another 
logic, which leads to the ``rooted formula'' 
each tree grows layer by layer from
a preferred root [AR1]. Our presentation here follows [AR1].

Applied to an element $H$ of  $\cS$, the
Taylor Rooted Forest formula takes the form
(the vector with all entries equal to $1$ being denoted by
$\bbbone$):

\medskip
\noindent{\bf Symmetric Taylor Forest Formula}{\it
$$
H(\bbbone)=
\sum_{\gF=\{l_1,\ldots,l_{\tau}\}
\atop{\rm u-forest}}
\biggl(\prod_{l\in\gF}\int_0^1 dw_l\biggr)
\biggl(\Bigl(\prod_{l\in\gF}
{\partial \over {\partial x_l}}\Bigr)H\biggr)
\bigl(X_\gF({\bf w})\bigr) \ \ .
\eqno({\rm 2.3})
$$
where the summation extends over all possible forests
$\gF$, including $\tau=0$ hence the empty forest. 
To each link of $\gF$ is attached a variable of integration $w_l$.
The vector $X_\gF({\bf w})$ is the value at which we evaluate the
derivative of $H$; it is the vector ${(x_l)}_{l\in\Pn}$
defined by $x_l=w_l^{\gF}({\bf w})$, 
where the weakening factor
$w_{\{ij\}}^{\gF}({\bf w})$ is 
$\inf \bigl\{w_l,l\in L_{\gF}\{ij\}\bigr\}$ if
$L_{\gF}\{ij\}$ is the unique path in the forest $\gF$ connecting i to j,
and is 0 if no such path exists
(hence if $i$ and $j$ belong to different clusters of $\gF$.}

The notation u-forest
stands for regular ``unordered'' forests. Decomposing the
integration domain into $\tau !$ subdomains according to a complete
ordering of the parameters $ w_{l}$ we obtain a related formula
where the sum runs over so called ``ordered'' forests in which the links
are ordered, and the integration parameters $ w_{l}$ follow the ordering. 

\medskip
\noindent {\bf Exercise 2.2} Write down the general ordered forest formula.
Remark that for $n=2$ it reduces to the ordinary Taylor formula
$H(1)= H(0) +\int_{0}^{1} xH'(wx) dw$. Performing a series of 
ordinary Taylor interpolations at each step restricted 
to the links joining different existing clusters, prove by induction
the ordered formula, from which the unordered one follows easily by regluing 
the integration domains for the $w$ parameters.$\hfill \clubsuit$\vspace{.15cm}

\noindent {\bf Exercise 2.3} Write down the ordered
and unordered formulas for 
$n=3$ and $4$ (very instructive). $\hfill \clubsuit$\vspace{.15cm}

Although we won't use it
let us stress that these formulas are not unique. For instance there exists
an absolutely identical formula, the rooted formula, with a different
rule for the weakening factor $w$, now called 
$w_{\{ij\}}^{\gF,R}({\bf w})$ (the superscript
$R$ standing for ``rooted'') . It is a less symmetric formula since we 
have to give a rule for choosing a root in each cluster.
For each non empty subset or cluster $C$ of $I_n$,
choose $r_C$, the least element in the natural ordering
of $I_n=\{1,\ldots,n\}$, to be the root of all the trees with support $C$
that appear in the following expansion.
Now if $i$ is in some tree $\gT$ with support $C$ we call the {\sl height}
of $i$ the number of links in the unique path of the tree $\gT$ that goes from
$i$ to the root $r_C$. We denote it by $l^\gT(i)$. The set of points $i$
with a fixed height $k$ is called the $k$-th {\sl layer} of the tree.
The Rooted Taylor Forest Formula is then absolutely identical
to the Symmetric one, except changing the superscript
$S$ to $R$ (as ``rooted'') and defining the weakening parameter
$w^R$ differently, by the following rule:

\medskip
\noindent{\bf Rooted weakening factors}
{\it

$w_{\{ij\}}^{\gF,R}({\bf w})=0$ if $i$ and $j$ are not connected by the $\gF$.
If $i$ and $j$ fall in the support $C$ of the same tree $\gT$ of $\gF$ then

$w_{\{ij\}}^{\gF,R}({\bf w})=0$\ \ if\ \ $|l^\gT(i)-l^\gT(j)|\ge2$\ \ 
($i$ and $j$ in distant layers)

$w_{\{ij\}}^{\gF,R}({\bf w})=1$\ \ if\ \ $l^\gT(i)=l^\gT(j)$\ \ 
($i$ and $j$ in the same
layer)

$w_{\{ij\}}^{\gF,R}({\bf w})=w_{\{ii'\}}$\ \ 
if\ \ $l^\gT(i)-1=l^\gT(j)=l^\gT(i')$,
and $\{ii'\}\in\gT$. ($i$ and $j$ in neighboring layers, $i'$ is then unique
and is called the ancestor of $i$ in $\gT$).
In particular, if $\{ij\}\in\gF$, then $w_{\{ij\}}^{\gF,R}({\bf w})
=w_{\{ij\}}$.}
\medskip

\noindent {\bf Exercise 2.4} Prove the rooted formula.
$\hfill \clubsuit$\vspace{.15cm}

It remains to see which formulas
can be used in the constructive sense and for which theories.
A very interesting property of the symmetric
formula is that it preserves positivity in a certain sense:

\medskip
\noindent{\bf Theorem 2 (Positivity of the Symmetric Formula)} 

{\it Extended to a symmetric matrix
by the convention $w_{ii}^{\gF}({\bf w}) =1 \forall i$,
the matrix $w_{ij}^{\gF}(\bf w)$ is positive.}
\medskip

\noindent {\bf Exercise 2.5} Prove this theorem (hint:
show that  $w_{ij}^{\gF}(\bf w)$ is a convex combination of block matrices
with 1 everywhere). $\hfill \clubsuit$\vspace{.15cm}

\subsection{Cluster Expansion}

Let us apply the previous formula
to study the thermodynamic limit of the {\it single slice} $\phi^4$
theory. The idea of the {\it cluster expansion} is that since
perturbation theory diverges we must keep most of it
in the form of functional integrals. However we can test whether
distant regions of space are joined or not by propagators, and
this will allow to rewrite the theory as a polymer gas (with hardcore 
interactions). When the coupling constant is small, the activities
for the polymers are small, and the technique of the Mayer expansion
which compares the hardcore gas to a perfect gas, allows to perform
the thermodynamic limit.

For instance  consider the free bosonic (massless) Gaussian measure 
$d\mu_{C}$ in $\RR^{d}$ defined by the single
slice covariance 
$$  C(x,y) = \int_{M^{-1}}^{1} 
{d\al \over \al^{d/2}} e^{-|x-y|^{2}/4\al}
\eqno({\rm 2.4})$$
This propagator is obviously integrable in $y$ at fixed $x$.

Put now the regular interaction $e^{-\lambda\int_{\La}\ph^{4}(x) } dx $
in a finite volume $\La$ and for simplicity let us study the pressure 
$$ 
p = \lim_{\La \to \RR^{d}}
{1\over | \Lambda |} \log Z(\Lambda) \ , \eqno({\rm 2.5})$$ 
where we recall that
the partition function $ Z(\Lambda)$ in a finite volume $\Lambda$ is 
$$ 
Z(\Lambda)= \int d\mu_{C}(\phi)e^{-\lambda\int_\Lambda\phi^4(x)dx} \ .
\eqno({\rm 2.6}) 
$$ 

Let us explain how the Taylor formula (2.3) performs the task
of rewriting the partition function as a dilute gas
of clusters with hard core interaction.

The set $I_{n}$ is defined as a partition of $\La$ into (unit size) cubes,
and clusters are subsets of such cubes.
We write $\La= \cup_{i\in I_{n}}b_{i}$, where each $b_{i}$ is a unit cube,
and define $\ch_{b}$ as the characteristic function of $b$, and $\ch_{\La}=
\sum_{i\in I_{n}} \ch_{b_{i}} $.
Since the interaction lies entirely within $\La$,
the covariance $C$ 
can be replaced by
$C_{\La}=  \ch_{\La}(x) C(x,y) \ch_{\La}(y)$ without changing the value
of $Z(\La)$. Moreover $C_{\La}$ can be 
interpolated, defining for $l=\{i,j\} \in \Pn$
$$
C_{\La}({(x_l)}_{l\in\Pn})(x,y) =  \sum_{i=1}^{n}\ch_{b_{i}}(x)  
C(x,y) \ch_{b_{i}}(y)$$
$$ + \sum_{\{i,j\}\in \Pn} x_{\{ij\}}
\bigl(\ch_{b_{i}}(x)  C(x,y) \ch_{b_{j}}(y)+
\ch_{b_{j}}(x)  C(x,y) \ch_{b_{i}}(y) \bigr)
\eqno({\rm 2.6})
$$
Remark that $C_{\La}(\bbbone ) = C_{\La}$.
Now we apply the  Taylor formula (2.3) with the function
$H$ being the partition function obtained by replacing in (2.6) 
the covariance $C$ by $C_{\La}({(x_l)}_{l\in\Pn})$.
Here it is crucial to use the positivity theorem, in order 
for the interpolated
covariance to remain positive, hence for the 
corresponding normalized Gaussian measure
to remain well defined. 

From the rules of Gaussian integration 
of polynomials [GJ], we can compute the effect of deriving
with respect to a given $x_{l}$ parameter, and we obtain
that (2.3) in this case takes the form
$$
Z(\La)= H(\bbbone) = 
\sum_{\gF}\int d\mu_{C_{\La}(X_{\gF}({\bf w})) }(\ph ) 
\biggl( \prod_{l\in \gF} \int_{0}^{1} dw_{l} \biggr)
$$
$$\biggl\{\prod_{l =\{ij\} \in \gF} \int dx dy \ch_{b_{i}}(x)
\ch_{b_{j}}(y)C(x,y)
 {\de \over \de \ph (x) }{\de \over \de \ph (y) }
\biggr\} e^{-\lambda\int_\Lambda\phi^4(x)dx}  \eqno({\rm 2.7})
$$
where $b_{i}$ and $b_{j}$ are the two ends of the line $l$.
Since both the local interaction
and the covariance as a matrix 
factorize over the  clusters of the forest $\gF$, 
the corresponding contributions in (2.7)
themselves factorize, which means that (2.7) can also be rewritten as
a gas of non-overlapping clusters, each of which has an amplitude
given by a {\it tree formula}: 
$$
Z(\La)= \int d\mu_{C_{\La} }(\ph )e^{I_{\La}(\ph)}=
 \sum_{{{\rm sets\ }\{Y_{1},...,Y_{n}\}} \atop
{Y_{i} \cap Y_{j} = \emptyset, \cup Y_{i} =\La}} \prod_{i=1}^{n} A(Y_{i})
\eqno({\rm 2.8})
$$
$$
A(Y) = 
\sum_{\gT \ {\rm on }\ Y} \bigl(
\prod_{l\in \gT}\int_{0}^{1} dw_{l} \bigr) 
\int d\mu_{C_{Y}(X_{\gT}({\bf w})) }(\ph )
$$
$$\biggl\{\prod_{l =\{ij\} \in \gT} \int dx dy \ch_{b_{i}}(x)
\ch_{b_{j}}(y)C(x,y)
 {\de \over \de \ph (x) }{\de \over \de \ph (y) }
\biggr\} e^{-\lambda\int_Y\phi^4(x)dx} 
\eqno({\rm 2.9})
$$
where the sum
is over trees $\gT$ which connect together the set $Y$, hence have exactly
$|Y|-1$ elements (if $|Y|=1$, $\gT=\emptyset$ connects $Y$). 
The measure $ d\mu_{C_{Y}(\{X_{\gT}({\bf h})\}) }(\ph)$ 
is the normalized Gaussian measure 
with (positive) covariance 
$$
C_{Y}(X_{\gT}({\bf w}))(x,y) = \ch_{Y}(x)\bigl(w_{\gT}
({\bf w})(x,y)\bigr)
C(x,y) \ch_{Y}(y)
\eqno({\rm 2.10})
$$
where $w_{\gT}({\bf w})(x,y)$ is 1 if $x$ and $y$ belong to the same cube,
and otherwise it 
is the infimum of the parameters $w_{l}$ for $l$ in the unique path
$L_{\gT}(b(x),b(y))$ which in the tree $\gT$
joins the cube $b(x)$ containing $x$ to the cube 
$b(y)$ containing $y$.

\medskip
\noindent {\bf Exercise 2.6} Prove that if $a_{ij}$ and $b_{ij}$
are two positive matrices, their Hadamard product  $c_{ij}=a_{ij}b_{ij}$
is again a positive matrix (hint: use square roots)
Complete the proof that $C_{Y}(X_{\gT}({\bf w}))(x,y)$ is, as announced,
a positive covariance, hence that it has a well defined associated Gaussian 
measure. $\hfill \clubsuit$\vspace{.15cm}

The following bound now summarizes that polymer activities are small
enough so that one can sum over all polymers containing a fixed point
(and absorb a fixed constant per cube of the polymer). This
will be used in the next section.

\medskip
\noindent{\bf Theorem 3 (Bound on Polymer Activities)} 

{\it  Given any constant $K$, for
small enough $\la$ with ${\rm Re} \ \la >0$ we have
$$
\sum_{Y \ {\rm such \ that}\ 0\in Y} |A(Y)| K^{|Y|} \le 1
\eqno({\rm 2.11})
$$}
\medskip

\noindent {\bf Exercise 2.7} Prove this theorem (hint:
use Formula (2.9). One needs notations to compute
the action of the functional derivatives in (2.9) by Leibniz rule,
the result being cumbersome. Then it is useful to remark that
the propagators exponential decay (see (2.4)) can absorb
any ``local factorial'' of the coordination numbers $d_{i}$ of the
tree $\gT$. This allows to achieve the proof
\footnote{This method is identical to the one in [R], part III.
But remark that although
the full amplitudes $A(Y)$ defined in (2.9) must be identical
to those in [R], the subcontributions associated to particular trees 
are different, since the formula used in [R] was not the symmetric one.}.
$\hfill \clubsuit$\vspace{.15cm}

\subsection{The Mayer expansion}

The Mayer expansion allows to deduce from (2.11) 
the existence and e.g. the Borel summability
in $\lambda$ of thermodynamic functions such as the pressure $p$.

In the cluster expansion (2.8), the condition
that the disjoint union of all clusters
is $\La$ is a global annoying constraint. 
Remark that the polymer 
amplitudes are translation invariant. In particular 
the trivial amplitude of a singleton cluster $Y=\{b\}$ is a number $A_{0}$
independent of $b$. Redefining $A_{r}(Y)= A(Y)/A_{0}^{|Y|}$
and $Z_{r}(\La) = Z(\La)/A_{0}^{|\La|}$ we quotient
out all the trivial clusters so that 
$$
Z_{r}(\La)= 1+
\sum_{n\ge 1} \sum_{{\rm sets }\ \{Y_{1},...,Y_{n}\} \atop
|Y_{i}| \ge 2 \ , \ Y_{i} \cap Y_{j} = \emptyset} 
\prod_{i=1}^{n} A_{r}(Y_{i})
\eqno({\rm 2.12})
$$
This is the partition function
of a polymer gas: the sums over individual polymers would be 
independent were it
not for the hard core constraints $ Y_{i} \cap Y_{j} = \emptyset$.
Adding in an infinite number of vanishing terms,
we can replace the sum in (2.12)
by a sum over ordered sequences $(Y_{1},...,Y_{n})$ of polymers
with hard core interaction and
a symmetrizing factor $1/n!$ coming from the replacement of sets by sequences.
$$
Z_{r}(\La)= 1+
\sum_{n\ge 1}  {1\over n!}\sum_{ {\rm sequences }\ (Y_{1},...,Y_{n}) \atop
|Y_{i}| \ge 2 \ } 
\prod_{i=1}^{n} A_{r}(Y_{i}) \prod_{1\le i< j\le n} \eta (Y_{i}, Y_{j})
\eqno({\rm 2.13})
$$
where the two-body hard core interaction is expressed by
the factors $\eta (X, Y)=1$ if $X \cap Y 
=\emptyset$, and $\eta (X, Y)=0$ if $X \cap Y 
\not =\emptyset$.
To factorize these hardcore interactions 
we apply again the symmetric Taylor forest 
formula (2.3)! More precisely for a fixed sequence $(Y_1,\ldots,Y_n)$ 
of polymers, we define $I_{n}$ as the set of these polymers, and 
define $\ep_{l}= \epsilon_{\{ij\}} = \eta (Y_{i}, Y_{j}) - 1$, for $i \ne j$.
We consider the function
$$
H({(x_l)}_{l\in\Pn}) = \prod_{l \in \Pn} (1 + x_l\ep_l) 
\eqno({\rm 2.14})
$$
so that $H(\bbbone)= \prod_{1\le i< j\le n} \eta (Y_{i}, Y_{j})$.
Rewrite (2.13) as
$$
Z_{r}(\La)= 1+
\sum_{n\ge 1}  {1\over n!}\sum_{ {\rm sequences }\ (Y_{1},...,Y_{n}) \atop
|Y_{i}| \ge 2 \ } 
 H(\bbbone ) \prod_{i=1}^{n} A_{r}(Y_{i}) 
$$
$$ = 1 + \sum_{n\ge 1}  {1 \over n!} 
\sum_{ {\rm sequences }\ (Y_{1},...,Y_{n}) \atop
|Y_{i}| \ge 2 \ } \prod_{i=1}^{n} A_{r}(Y_{i}) \sum_{\gF} 
\biggl( \prod_{l\in \gF} \int_{0}^{1} dw_{l} \biggr)
\biggl( \prod_{l\in \gF} \ep_{l} \biggr) 
\biggl( \prod_{l\not\in \gF} (1+ w_{l}^{\gF} ({\bf w}) \ep_{l}) \biggr) 
$$
$$=  \sum_{n\ge 0}  {1 \over n!}\biggl( 
\sum_{k\ge 1}  {1 \over k!}\sum_{ {\rm sequences }\ (Y_{1},...,Y_{k}) \atop
|Y_{i}| \ge 2 \ } \Bigl(\prod_{i=1}^{k} A_{r}(Y_{i})\Bigr)
C^{T}(Y_{1},...,Y_{k})  
\biggr)^{n}
\eqno({\rm 2.15})
$$
where
$$
C^{T}(Y_{1},...,Y_{k}) = 
\sum_{\gT \ {\rm tree\ on \ }  \{1,....,k\}} 
\biggl( \prod_{l\in \gT} \int_{0}^{1 } dw_{l}\biggr)
\biggl(\prod_{l\in {\cal P}_{k}, \ l\in \gT}  \ep_{l}\biggr)
\prod_{l\in {\cal P}_{k}, \ l \not \in \gT} \bigl(1 + w_{l}^{\gT}({\bf w}) 
\ep_{l} \bigr)
\eqno({\rm 2.16})
$$
where, as before, $ w_{l}^{\gT}({\bf w}) $ is, if $l=\{ij\}$, the  
infimum of the parameters $w_{l'}$ for $l'$ in the unique path
$L_{\gT}\{ij\}$ which in the tree $\gT$
joins $i $ to $j$.

\medskip
\noindent {\bf Exercise 2.8} Check that the connecting factor
$C^{T}(Y_{1},...,Y_{k})$ does not depend of the particular tree
formula, since we have
$$C^{T}(Y_{1},...,Y_{k}) = 
\sum_{{G \ {\rm connected\ graph}}\atop{{\rm on} \{1,....,k\}}}
\prod_{l\in G} \ep_{l} .\eqno({\rm 2.17})
$$
$\hfill \clubsuit$\vspace{.15cm}

We obtain immediately that 
$$
\log Z_{r}(\La) = 
\sum_{k\ge 1}  {1 \over k!}\sum_{ {\rm sequences }\ (Y_{1},...,Y_{k}) \atop
|Y_{i}| \ge 2 \ } \biggl(\prod_{i=1}^{k} A_{r}(Y_{i})\biggr)
C^{T}(Y_{1},...,Y_{k})
\eqno({\rm 2.18})
$$

These formulas can be used together with (2.11)
to control the thermodynamic limit
$p = A_{0} + \lim _{\La \to \RR^{d}} {1 \over |\La |}
\log  Z_{r}(\La)$ (hint [R]: 
Every tree coefficient forces the necessary overlaps
along the links of the tree,
and is bounded by 1, since $| (1 + w_{l}^{\gT}({\bf w}) \ep_{l})|\le 1$.
Start from the $d_{i}-1$ leaves of the tree hooked to a given polymer
of index $i$, 
fix the $d_{i}-1$ squares of overlap of these leaves with the polymer
(at a cost $|Y_{i}|^{d_{i}-1}$), then sum over the leaves using the 
decay of (2.11) in the
size of polymers. Use the factor 
${1\over (d_{i}-1)!}$ in Cayley's theorem (Exercise 2.1) and the fact
that $\sum_{d_{i}} |Y_{i}|^{d_{i}-1}/(d_{i}-1)!
= e^{ |Y_{i}|}$. Then absorb the factor $e^{ |Y_{i}|}$ in the constant
$K$ of (2.11). Finally iterate to conclude, fixing the last sum to contain
the origin, since this cancels out the volume factor $|\La |$ 
in (2.5), up to boundary terms which vanish as $|\La |\to \RR^{d}$.).

\medskip
\noindent {\bf Exercise 2.9} Prove Borel summability
of the theory in a single slice, using the Nevanlinna theorem
of P.da Veiga's lectures (hint: the analyticity for 
${\rm Re} \la > 0$ and $\la$ small is easy. For the $K^{n} n!$ bounds
on the Taylor remainder at order $n$, simply rewrite the same analysis
than above but
with $n$ explicit vertices expanded first.
$\hfill \clubsuit$\vspace{.15cm}

A full renormalizable model such as the infrared critical
point of $\ph^{4}_{4}$ can be built by extending this single slice
analysis to the full multislice model. Let us sketch very briefly how
this model would be treated in the multislice expansion
of [AR2]. The analysis decomposes into three main steps.

The multislice model
contains propagators $C^{j}$ in each momentum slice, and space has
to be therefore decomposed at each slice into a scaled lattice
$(M^{j}\ZZ)^{4}$. The union  $\cup_{j} (M^{j}\ZZ)^{4}$ is the full
phase space for the theory, each cube being some kind
of degree of freedom for the appropriate field $\phi^{j}$. The
first expansion step is a ``small versus large field decomposition'' 
which tests, for each cube $b\in \cup_{j}(M^{j}\ZZ)^{4}$ of scale $j(b)$,
whether a quantity such as  $\int_{b} (\phi^{j(b)})^{4}$ is small or not.
The maximal connected regions of large field cubes will be treated as single
blocks, i.e. no cluster expansion will be performed inside them. This step,
although perhaps not strictly necessary, simplifies a lot the bounds later.

The second step is the multiscale ``cluster expansion''. 
It has to decouple now all the small field cubes and all the large 
field blocks of all slices $j=0,...\rho$ in (1.10) at the same time.
This means that it derived both propagators $C^{j}$ in each momentum slice, 
this time joining cubes of scaled lattices $(M^{j}\ZZ)^{4}$. But it has
also to derive new links which join the
cubes of different slices at a same point, 
this time through vertices. These new ``vertex links''
can join up to 4 cubes at the same time, since a vertex can hook
to at most 4 propagators in different slices. Therefore the symmetric
Taylor formula has to be properly extended to accommodate
such new links [AR2]. Also the formulas in the ``momentum slice''
direction have to be pushed further than simple connectedness,
since we want to distinguish convergent activities
(those with more than 5 external lower momentum legs)
from divergent ones (with 2 and 4 external lower momentum legs),
which require renormalization. As a result the formula in the 
slice (or ``vertical'') direction is not a tree formula,
but rather some kind of a ``5 particle irreducible'' formula,
with an expansion rule which stops the vertical or vertex interpolations only 
when 5 or more links are derived. In spite of these complications,
at the end of the multiscale cluster expansion
the theory is factorized as a gas of polymers swimming
in the full phase space, hence in $\cup_{j}(M^{j}\ZZ)^{4}$, with hardcore 
interactions.

The third step is the generalized Mayer expansion, whose role is no longer 
only to factorize and quotient out the normalization (vacuum graphs), but also
to factorize and renormalize the 2 and 4 point functions. Indeed
only the convergent polymers, those which do not contain subpolymers with 2
and 4 external lower momentum legs, of the previous expansion have small 
summable activities. Divergent polymers require renormalization,
the appropriate counterterms being absorbed into effective constants.
This generates more complicated formulas, but the logic remains 
the same (interpolate the hardcore links). The result is some
complicated but totally explicit formula for expanding the model. It
can be viewed as the correct constructive analog
of Zimmermann's forest formula for perturbative renormalization [A]. 

This method, although still very heavy and very technical, is nevertheless
clearly more explicit and conceptually more transparent than previous methods 
in which cluster and Mayer expansion steps were mixed together in an 
inductive way (such as in [FMRS2]).

\section{Fermionic theories}

\subsection{Introduction}

The initial constructions of renormalizable
Fermionic models such as the Gross-Neveu$_{2}$ model
[GK1] or [FMRS1] used the same heavy apparatus as Bosonic models, namely
slicings in momentum space and direct space, plus cluster and
Mayer expansions. As a consequence the renormalization group equations 
obtained were discrete difference equations instead of differential equations,
a detail which was considered annoying by some physicists. 
On the other hand 
perturbation theory for Fermion systems is often said to converge,
whereas for boson systems it is said to diverge. But what does this mean 
exactly? Unnormalized Fermionic perturbation series with cutoffs
are not only convergent but {\it entire}, whereas Bosonic perturbation 
series, even unnormalized and with cutoffs, have zero radius of convergence. 
Using this property there ought to be constructive versions
of renormalization for Fermions very close to the perturbative concepts. 
In particular it was advocated recently in [S] 
that there ought to exist continuous flows and differential
equations of the renormalization group for Fermions.

Progressively we realized that one can avoid the use
of the cluster and Mayer expansions of the previous section 
for Fermionic models (for an early example see [FMRT1]). In
particular the symmetric forest formula of the previous section,
when the interpolated parameters are directly applied
to the Feynman lines of a graph, gives a particularly simple
``three lines'' construction of interacting Fermions in a single
momentum slice [AR3], which is given below. In [DR] we applied similar 
ideas to give a new construction of the Gross-Neveu$_{2}$ model. This 
construction gives a
solution of the theory in the form of explicit sums of
finite dimensional integrals, containing, however,
effective constants which are defined as the (non explicit) solutions
of the differential renormalization group equations.

The positivity property of the symmetric Taylor forest
formula (Theorem 3 of the previous section) was crucial for 
Bosonic models. But here again it is useful, since
it allows to apply Gram's estimates:

\medskip
\noindent{\bf Lemma 1 (Interpolated Gram inequality)}
\medskip
{\it
Let $A=a_{ij}=<f_{i}\cdot g_{j}>$ be a Gram matrix. Gram's inequality
remains true for the matrix $B=b_{ij}=  w_{ij}^{\gF} ({\bf w})
<f_{i}\cdot g_{j}>$, namely:
$$ |det B| \le \prod_{i} || f_{i}||\prod_{j} || g_{j}|| \quad
\quad\quad  \ \forall \bf w \eqno({\rm 3.4})
$$}

\prf  Indeed for fixed $\gF$  and $({\bf w})$
we can take the symmetric square root $v$ of the 
positive matrix $w^{\gF}({\bf w})$, 
so that $w_{ij} = \sum_{k} v_{ik}v_{kj} $. Defining
the components of the vectors $f$ and $g$ in an orthonormal basis
for the scalar product $< \cdot >$ to be $f^{m}_{i}$ or $g^{m}_{j}$,
we define the tensorized vectors $F_{i}$ and $G_{j}$ with components
$F_{i}^{km}=v_{ik}f_{i}^{m}$ and $G_{j}^{km}=v_{jk}g_{j}^{m}$ and
we have for the tensor scalar product $<\cdot >_{T}$:
$b_{ij} = < F_{i} \cdot G_{j}>_{T}$, so that $det B \le
\prod_{i} || F_{i}||_{T}\prod_{j} || G_{j}||_{T} $. But obviously
$||F_{i}||_{T}^{2}= \sum_{km}( F_{i}^{km})^{2} = \sum_{km} v_{ik}^{2} 
(f_{i}^{m})^{2} = w_{ii} \sum _{m}(f_{i}^{m})^{2} =||f_{i}||^{2} $
(we recall that by definition $ w_{ii}^{\gF} ({\bf w})=1$).
\qed
\medskip

Let us now apply the Symmetric Taylor Forest formula 
to the computation of the connected functions of a Fermionic theory.
The corresponding Grassmann integral is:

$${1\over Z}\int d\mu_{C}(\ps, \bar \ps )
P(\bar\psi_{a}, \psi_{a} )e^{S(\bar\ps_{a}, \ps_{a})}
\eqno({\rm 3.5})
$$
where $C$ is the covariance or propagator,
$P$ is a particular monomial (set of external fields)
and $S$ is some general action. We take as simplest example
the massive Gross-Neveu model with cutoff, for which the 
action is local and quartic in a certain number of 
Fermionic fields. 
In a finite box $\Lambda$ the action is
$$S_{\Lambda}= {\lambda \over N}\int_{\Lambda} dx 
(\sum_{a}\bar\ps_{a}(x)\ps_{a}(x))^{2}\eqno({\rm 3.6})
$$
where $a$ runs over some finite
set of $N$ ``color'' indices. 
$\lambda$ is the  coupling constant. There may be also a spinor index
which we forget in our notations
since it plays no role in what follows.
The covariance $C$ is massive and has an ultraviolet
cutoff, hence in Fourier space it is for instance
$\et(p)/(\not p +m)$  where $\et$ is a cutoff function on large momenta.
We only need to know that $C$ is diagonal in color space
$ C(x, a; y, a') = 0 $ for $a \ne a'$, and that it can be decomposed
as 
$C(x,y)= \int_{\R^{d}} D(x,t) \bar E(t,y)dt$
with
$$ |C(x,y) |\le K {1 \over (1 + |x-y|)^{p}} \eqno({\rm 3.7a})
$$
$$ \int_{\R^{d}} |D(x,t)|^{2}d^{d}t \le K \quad ;
\quad \int_{\R^{d}} |E(x,t)|^{2} d^{d}t\le K \eqno({\rm 3.7b})
$$
for some constant $K$.

This decomposition amounts
roughly to defining square roots of the covariance in momentum
space and prove their square integrability. It is usually  
easy for any reasonable cutoff model. For instance if $\et$ 
is a positive function, we can define $D$ in Fourier space
as ${\et^{1/2}(p) \over (p^{2} + m^{2})^{1/4}}$ 
and $E$ as ${(-\not p + m) \et^{1/2}(p) \over (p^{2} + m^{2})^{3/4}}$.

It is the Fermionic covariance or propagator $C= <\ps_{a}(x)
\bar\ps_{a}(y)>$ which is interpolated with the forest formulas. 
We obtain, for instance
for the pressure that the formal power series in the coupling constant is
a sum over trees on $\{1,....,n\}$, with a distinguished
vertex, 1, sitting at the origin to break translation invariance
(similar formulas with external fields of course
hold for the connected functions):

\subsection{Fermionic Tree Expansion}
{\it
$$
p= \lim_{\Lambda \to \infty} {1\over |\Lambda|} Log Z(\Lambda) =  
\lim_{\Lambda \to \infty} {1\over |\Lambda|}\bigl(
\int d\mu_{C }(\ps, \bar \ps )e^{S_{\Lambda}(\bar\ps_{a}, \ps_{a})}\bigr)
$$
$$
= \sum_{n=0}^{\infty}(\la^{n}/N^{n}n!)
\sum_{\gT,{\bf \si}} \sum_{{\cal C}}  \ep(\gT,{\bf \si}) \bigl(
\prod_{l\in \gT}\int_{0}^{1} dw_{l} \bigr) 
\int dx_{1}...dx_{n} \de(x_{1}=0)
$$
$$
\biggl\{\prod_{l =\{ij\} \in \gT} C_{\si(l)}(x_{i},y_{j})
\biggr\}
\int d\mu_{C_{X_{\gT}({\bf w})} }(\ps, \bar \ps ) 
\prod_{r=1}^{n+1} \bar \ps (x_{i(r)}, a(r)) \ps (x_{j(r)} a'(r)) 
\eqno({\rm 3.8})
$$
where the sum
is over trees $\gT$ which connect together the $n$ points 
$x_{1},...,x_{n}$. These trees also contain an arrow information
${\bf \si}$ which for each line determines which end 
was a field and which was an antifield. The coloring
${\cal C}$ contains the color indices of each line of the tree and
of each remaining field or antifield.
This information completely determines the remaining set 
of uncontracted fields $\prod_{r=1}^{n+1} \bar 
\ps (x_{i(r), a(r)}) \ps (x_{j(r), a'(r)})$,
that is it determines the functions $i(r)$ and $j(r)$
which tell to which vertex the remaining fields are hooked, and the
coloring index ${\cal C}$ completely determines the functions $a(r)$, 
$a'(r)$ (hence their color). Finally
$\ep(\gT,{\bf \si})$ is some sign $\pm 1$ which we do not need to compute.}

\medskip
\noindent {\bf Exercise 3.1} Find first a similar but simpler
formula for the number of connected graphs of the $\phi^{4}$ 
theory at order $n$ as a sum over
trees with weakening factors. (Counting graphs with
their correct combinatoric factor is equivalent to field theory
in zero dimension). Check explicitly in the cases $n=3$ and  $n=4$
how the weakening factors when integrated restore the right combinatorics
for the ordinary graphs (very instructive). $\hfill \clubsuit$\vspace{.15cm}

\noindent {\bf Exercise 3.2} Check that (3.8) is indeed the result
of applying the Symmetric Taylor Forest formula to the Grassmann
functional integral (3.5), interpolating the Fermionic
covariance $C$ and using the Grassmannian 
rules of ``integration by parts''. 
For the courageous reader: find an explicit
formula for the sign $\ep(\gT,{\bf \si})$.
$\hfill \clubsuit$\vspace{.15cm}

\medskip
\noindent{\bf Lemma 2}
{\it The index $\cal C$ in the sum (3.8) runs over 
a set of exactly $2^{n}N^{n+1}$
elements.}

\prf At each vertex the circulation of color indices is fixed
by a factor 2 (which tells which of the two fields $\psi$ is paired
with a $\bar \psi$ i.e. forced to have the same color by the form (3.6)
of the action. Once these circulation rules are fixed,
the attribution of color indices costs $N^{2}$ for the first vertex,
and by induction a factor $N$  for each vertex of the tree.
Indeed climbing inductively into the tree layer by layer, at every vertex
there is one color already fixed by the line joining the vertex 
to the root, hence one remaining color to fix, except for the root, 
for which two colors have to be fixed. This proves the lemma. \qed

\medskip
Then the expression $$\int d\mu_{C_{X_{\gT}
({\bf w})} }(\ps, \bar \ps ) 
\prod_{r=1}^{n+1} \bar \ps (x_{i(r), a(r)}) \ps (x_{j(r), a'(r)})
\eqno({\rm 3.9}) 
$$
is nothing but an $n+1$ by $n+1$ determinant, with matrix element
the covariance 
$$w_{l}^{\gT}({\bf w})\de(a(r),a'(r'))
C(x_{i(r)},x_{j(r')})\eqno({\rm 3.10}) $$ 
between the line $r$ and the column
$r'$, where $l=(i(r),j(r'))$. 

Let us now use these Fermionic tree formulas for proofs of convergence.

\subsection{Convergence of the 
tree formulas}

A typical constructive result for this Gross-Neveu model with cutoff 
is to prove:

\medskip
\noindent{\bf Theorem 4}
{\it The pressure and the connected functions of the cut-off Gross-Neveu
model are analytic in $\lambda$ in a disk of radius $R$
independent of $N$.
}
\medskip

\prf Apply Gram's inequality to the  loop lines in (3.8).
By Lemma I and (3.7b), this determinant is bounded by $K^{n+1}$.
The spatial integrals are bounded, using (3.7a) by $K^{n-1}$.
The number of trees on $n$ vertices is $n^{n-2}$ by Cayley's
theorem. In the sum over colors, the coloring factor
$N^{n+1}$ of Lemma 2 almost cancels
with the factor $1/N^{n}$ in (3.8).
Hence the power series in $\la$ for the pressure is
 bounded by $ N {n^{n-2}\over n!}{ K'}^{n}$ for some constant $K'$. \qed

\medskip
This is perhaps the shortest and most transparent proof
of constructive theory yet!\footnote{One can also prove 
the same result using the rooted formula rather than the symmetric one [AR3]. 
Although the proof is somewhat longer, it is
interesting since the weakening factor
in the rooted formula completely factorizes out of the determinant.
This second formula may therefore be useful 
in problems for which Gram's inequality is not 
applicable and the method of ``comparison of rows and columns''
of [IM2] or [FMRT1] has to be used. This method
roughly corresponds to Taylor expanding around
a middle point further and further when fields
or antifields accumulate in any given cube of unit size
of a lattice covering $\R^{d}$.} 
The theorem above is interesting not only 
for the analysis of the Gross-Neveu
model but also for that of the two-dimensional
interacting Fermions considered in the next section. In this latter case,
the ``colors'' correspond to angular sectors on the Fermi
sphere and the factor $1/N$ in the coupling is provided by power counting.

\subsection{Renormalization, an overview}

In this subsection we give a brief summary of how to remove
the ultraviolet cutoff $\eta$ and perform renormalization of
the Gross-Neveu$_{2}$ model, following [DR].

The covariance $C$ in Fourier space with infrared cutoff
$\Lambda$ (this is no longer the volume!) and ultraviolet cutoff 
$\Lambda_{0}$ can be written as
$$ C_{\La}^{\La_{0}} = 
 {(- \not p + m  )\over  p^{2} + m ^{2} }
(e^{-\Lambda^{-2} (p^{2}+m^{2})} - e^{-\Lambda_{0}^{-2} (p^{2}+m^{2})} )
=  \int _{\Lambda_{0}^{-2}} ^{\Lambda^{-2}} (- \not p + m  )
e^{-\al (p^{2}+m^{2}} \eqno({\rm 3.11})$$

Our goal is to prove that the vertex functions 
of the theory have a non perturbative limit as $\Lambda_{0}\to \infty$,
and that they are the Borel sum of their renormalized power series
in the renormalized coupling constant (because of renormalons they cannot
be analytic).

The main idea is to apply the interpolation forest formula not directly
to the lines of the Feynman graphs, but to the continuous scale parameter
$\al$ introduced in (3.11). In this way an {\it ordered} forest formula
is built, in which the ordering of the tree lines plays the role of the
necessary ordering of momenta in any phase space analysis, namely
to distinguish higher momenta from lower momenta. This expansion
alone however, does not put explicitly into
display the divergent subgraphs with two and four external legs.
An additional construction (classes of ``chains" in [DR]) 
has to be performed, to expand
further the remaining loop determinant in (3.8) so that this structure
becomes visible. This remaining expansion has to be performed with some
care. Indeed we cannot simply sum over the attributions of 
the $\al$ parameters of the loop lines to the bands defined
by the intervals left between the
$\al$ parameters of the tree lines. This would make visible
not only the 2 and 4 point high energy subgraphs but also all the
other high energy subgraphs, with arbitrary number of external legs.
And it is a general rule of this kind of mathematical problems that information
has a price. This particular information would lead to uncontrolled divergences
at large order. The solution found in [DR] is to reglue together many
band attributions into so-called ``chain classes'' so as to expand
much less the loop determinant. After this regluing, the only information
on a high energy subgraph displayed by the expansion is whether this subgraph
has  2, 4, or more than 4 low energy legs, and {\it this}
information alone does not introduce large order divergences.

After this subtle point is settled, the rest of the analysis
is completely similar to perturbative renormalization theory.
Counterterms for 0 momentum values are introduced which subtract
the two and four point high energy subgraphs and create flows  
for the three relevant or marginal operators
of this theory, namely the coupling constant, the wave function constant and
the mass. Technically the subtractions are performed in direct $x$ space,
since again we do not want to know the exact structure of loop lines,
which would be necessary for Zimmermann's subtraction 
scheme in momentum space.

Remark that the initial 
ordering of the tree lines in this formalism
reminds of the ordering of Hepp's sectors in perturbative
renormalization theory (see [R], and references
therein); but it orders only about ``half'' the lines
of a graph (namely the tree lines) whereas Hepp's sectors ordered
all the lines of a graph.

The formalism has also some similarities
to the Gallavotti-Nicol{\`o} tree expansion [GN], but beware that the 
basic trees in [DR] are {\it not} the Gallavotti-Nicol{\`o}
trees, which correspond rather to the auxiliary
trees considered for the construction of the ``chain classes''.

\section{Renormalization Group in Condensed Matter}

\subsection{Many Interacting Fermions and the BCS Problem}

Conducting electrons in a metal at low temperature are well described
by Fermi liquid theory. However we know that the Fermi liquid theory is
not valid until temperature 0. Indeed below the BCS critical temperature
the dressed electrons or holes which are the excitations of the Fermi liquid
are bound into Cooper pairs and the metal becomes superconducting.

During the last ten years a program has been designed to investigate
rigorously this phenomenon by means of field theoretic methods 
[BG][FT1-2][FMRT1][S]. In particular the renormalization group of Wilson 
and followers has been extended to models with surface singularities such as 
the Fermi surface. The ultimate goal is to create a mathematically rigorous 
theory of the BCS transition and of similar phenomena of solid state physics. 
From the start we know that nonperturbative effects must be incorporated in 
the analysis of superconductivity since the BCS gap is nonperturbative. 
But this model is the only one we know of in which 
nonperturbative effects should be accessible in the near future to 
rigorous mathematical control, without any ad hoc modifications 
(such as introducing an artificially large number of colors). Indeed
an amazing property of this model 
is that angular variables around the Fermi surface play a role
analogous to that of colors, so that an expansion of the $1/N$ 
type should control the BCS regime [FMRT2], in which ordinary perturbation
is no longer valid. We can call this situation a ``dynamical $1/N$''effect.

Nevertheless the full construction of the BCS model
is a long and difficult program which requires to glue together several
ingredients. The main new idea was to extend
the renormalization group of Wison (which analyzes the point singularity
$p=0$ in momentum space) to more general extended singularities. This very 
natural and general idea is susceptible of many applications in various 
domains, including field theory (in Minkowski space). But this idea has also 
to be combined with rigorous control of spontaneous continuous symmetry 
breaking, and this generates a lot of technical complications.
In these lectures we therefore restrict us to the study of
the concept of RG around the Fermi surface in a simpler
model which is at a temperature higher than the temperature
of the BCS phase transition.

The key feature which differentiates electrons in condensed matter from 
Euclidean field
theory, and makes the subject in a way mathematically richer, is that Lorentz
invariance is broken, and density is finite. The field theory formalism
remains the best tool to isolate the fundamental
issues such as the existence of non-perturbative effects.
Imaginary (Euclidean) time (in the form of a circle,
with antiperiodic b.c. for Fermions) corresponds to finite temperature.
The fundamental state of the theory corresponds to the limit
of the temperature going to 0.
In this limit the imaginary time circle grows to $\RR$.
Finite density creates the Fermi sea and surface. In the simplest
case where rotation invariance is preserved (isotropic jellium) this surface
is simply a sphere.

The free Fermi liquid theory is therefore
defined by Fermion fields with two spin indices, and propagator
$$
C_{ab} (k) = \de_{ab} \frac{1}{ik_0-e(\vec{k})}
\quad; \quad e(\vec{k})= \frac{\vec{k}^2}{2m}-\mu
\eqno({\rm 4.1})
$$
where $a,b \in \{1,2\}$ are the
spin indices. The vector $\vec k$ is in $d$ spatial dimensions. 
Adding one time dimension there are really $d+1$ dimensions.
The parameters $m$ and $\mu$ correspond to the effective mass and 
the chemical potential (which fixes the Fermi energy).  
To simplify notation we put $2m=1$, $\mu=1$ so 
$e(\vec{k})= \vec{k}^2-1$.

This propagator is rotation invariant, a feature which simplifies 
considerably the study of the renormalization group flows
after branching the interaction. In particular it
has a spherical Fermi surface. This jellium isotropic model
is realistic for instance in solid state physics
in the limit of weak electron densities (where the Fermi surface becomes
approximately spherical), but in general a propagator with a more
complicated energy function $e(\vec{k})$ (such as a lattice Laplacian)
has to be considered. This is not necessary for
our purpose here which is to test the constructive validity of
perturbation theory.

Since we work at finite temperature and since
Fermionic fields have to satisfy antiperiodic boundary
conditions, the component $k_0$ can take only discrete
values (called the Matsubara frequencies) :
$$
k_0 = \pm \frac{2n+1}{\beta  \hbar
} \pi\eqno({\rm 4.2})
$$
so the integral over $k_0$ is really a discrete sum over $n$. 
As $k_0\neq0$ $\forall n$ the denominator in $C(k)$ can never be 0.
This is why the temperature provides a natural infrared cut-off. 
When $T \to 0$ $k_0$ 
becomes a continuous variable and the propagator diverges on the Fermi
surface, defined by $k_0=0$ and $|\vec{k}|=1$. 

The interaction term in the action is defined by:
$$
S_{\Lambda} = \frac{\la}{2} \int_{\Lambda} d^3x\; (\sum_a \bar \psi\psi)^2
\eqno({\rm 4.3})
$$
Physically this interaction may represent the effective
interaction due to phonons. A more relaistic interaction
would not be completely local to include the short range nature
of the phonon propagator, but we can 
consider the local action (4.3) as an idealization which captures all
essential mathematical difficulties.

\subsection{Renormalization around the Fermi Surface}

It is convenient to add a continuous ultraviolet cut-off 
(at a fixed scale) to the propagator
(4.1) because it makes its Fourier transformed
kernel in position space well defined, and because a non relativistic
theory does not make sense anyway at high energies.

The basic difference between this theory and an ordinary
critical point in statistical mechanics or
field theory, is that the singularity of the propagator
is of codimension 2 in the $d+1$ dimensional space-time.
This changes dramatically the power counting. Instead of changing
with dimension, power counting in this kind of models
is essentially independent of the dimension, and is the
one of a just renormalizable theory. This can be understood
basically in the following way. In a graph with 4 external legs, there are
$n$ vertices, $2n-2$ internal lines and $L=n-1$ independent loops. 
Each independent loop momentum gives rise to two transverse variables
$k_{0}$ and $|\vec k |$ and $d-1$ inessential bounded angles.
Hence the $2L=2(n-1)$ dimensions of integration for
the loop momenta exactly balance
the $2n-2$ singularities of the internal propagators,
as is the case in a just renormalizable theory. 

To justify this very crude picture, it turns out that
it is very convenient to further decompose
the propagator into discrete slices and each slice into discrete
angular sectors\footnote{A continuous slicing
in the style of [DR] is also possible;
but the discrete slicing is more in Wilson's spirit.}:
$$
C= \sum_{j=1}^{\infty}C^{j}
\quad ; \quad C^{j} (k) = 
{f^j(k) \over ik_0-e(\vec k)}\eqno({\rm 4.4})
$$ where 
$$ f^j(k )=f\left(M^{2j}\left(k_0^2+e(\vec k)^2\right)\right) 
\eqno({\rm 4.4})$$
effectively forces $|ik_0-e(\vec k)|\sim M^{-j}$.  The function $f$ is in
$C^\infty_0([1,M^{4}])$. The parameter $M$ is strictly bigger than one so
that the slices pinch more and more the Fermi surface as $j\to \infty$.

The slice propagator is further decomposed into sectors:
$$C^{(j)} (k) = \sum_{\si \in \Si^{j}}  C^{j, \si} (k)
\quad ; \quad C^{j,\si} (k) = {f^{j,\si}(k)
\over ik_0-e(\vec k)}\eqno({\rm 4.5})
$$
where $\Si^{j}$ is a set of roughly $M^{(d-1) j}$ angular patches,
called sectors, which cover the Fermi sphere, with linear dimensions
of order $M^{-j}$. For instance if $d=2$ we simply cut the circle
into intervals of length $2 \pi M^{-j}$.

The RG philosophy applied to this problem is now clear. As before,
higher slices give rise to local effects relatively to
lower slices. After more and more slice integrations one obtains effective
actions which govern longer and longer distance physics.
These effective actions are however more 
complicated than in the field theory context.
In rotation invariant models such as the one above, renormalization
of the two point function can be absorbed in a change
of normalization of the Fermi radius. It removes all infinities
from perturbation theory at generic momenta [FT1]. 
But the flow for the four point function is
a flow for an infinite set of coupling constants describing
the momentum zero channel of the Cooper pairs [FT2]. In the
case of attractive interaction for $\la$, if the temperature
cutoff is lowered to zero, this flow diverges
at the BCS scale, where the symmetry linked to particle number
conservation is spontaneously broken, giving rise to the
effective BCS theory for the Cooper pairs.

Like in the previous section the key problem from the
constructive point of view
is to understand the resummation of perturbation theory
in a single slice, in order to iterate
the renormalization group step.  

Curiously, although power counting does not depend on the
dimension, momentum conservation in terms of sectors 
in a fixed slice depends on it
and this has crucial constructive consequences. 
In $d=2$ we have the ``rhombus rule'':
four sectors meeting at a vertex must be roughly two by two
equal\footnote{A precise version is given in [FMRT1]. Remark that
there is a logarithmic correction to this rhombus rule, due to
the case of a nearly collapsed rhombus. A nice 
improvement to avoid further problems with this logarithm
is to define angular sectors longer in the tangential
than in the radial direction.}.
This means that two dimensional condensed matter in a slice
is directly analogous to an $N$-vector model in which angles 
on the Fermi surface play the role of
colors [FMRT3]. This allows to complete the proof that the
radius of convergence of perturbation theory is
independent of the slice index $j$.
Roughly speaking, the model in slice
$j$ is a vector model with $N=M^{j}$
colors. Power counting at a vertex costs $M^{3j}$ for
the space-time integration, and the propagators in fixed sectors
earn a scaling factor $M^{-2j}$ each. Since there are in average
two propagators per vertex, we remain with a factor $M^{-j}=1/N$ left 
per vertex, which is exactly what is necessary to pay for the color sums 
and obtain a uniform radius of convergence by Theorem 4 of section 3.
This completes the sketch of the proof of:

\medskip
\noindent{\bf Theorem 5 [FMRT1]}
{\it \  In two dimensions, there exists a finite $\kappa>\ 0\ \ $ 
independent of $j$ such that the power series in $\lambda$ 
of the Schwinger function for the interacting Fermionic measure
with propagator (4.5) and interaction (4.3)
has a convergence radius of at least $\ka $.}

\medskip

Furthermore it ought to be possible in two
dimensions to resum perturbation theory to build an effective action
until the BCS temperature. As an example, the convergent
part of the expansion (which does not include
2 or 4 point subgraphs) has been resummed in [FMRT1].

In three dimensions, two momenta at a vertex in a slice do not
determine the third and fourth: there is an additional torsion angle,
since four momenta of same length adding to 0 are not necessarily
coplanar. The radius of convergence of perturbation theory is still
independent of the slice index $j$, so that Theorem
5 also holds in three dimension  [MR], but it is much harder to
prove than in dimension 2, and until now it is not clear
that this partial result alone allows a full constructive analysis
of the model up to the scale where the BCS symmetry breaking takes place.

In dimension 1 the Fermi surface reduces to two points,
and there is also no proper BCS theory since there is no
continuous symmetry breaking in two dimensions (by the 
``Mermin-Wagner theorem'').
Nevertheless the many Fermion system in 1 dimension gives
rise to an interesting non-trivial behavior, that of the Luttinger
liquid [BG].

\subsection{The Weakly Coupled Anderson Model} 

This model describes a {\it single} electron in a random
potential. This is not strictly speaking
a field theory. But renormalization group
and field theoretic methods can also be applied to
models such as self repelling walks [IM1]. This model
lies in this category. 
The two-point function is given by
$$
{\cal S}(x,y ; E ; \lambda V ; \epsilon)\ =\ 
\Big[{1\over -\Delta-E\ +\ \eta \lambda V\ +\ i \epsilon}\Big](x,y)
$$
$$
=\ \Big[\big({1\over p^{2}-E\ +\ i \epsilon}\big)\ \     {1\over 
1\ +\ {\lambda\over p^{2}-E +\ i \epsilon}  \eta  V}   \Big](x,y)
\eqno({\rm 4.5})
$$
where $\ \eta\ $ is an ultraviolet cutoff and $\ V\ $ is a random
potential (multiplication operator in $x$-space),
for instance distributed with a Gaussian (regularized) white noise
for which the covariance is a (regularized) delta function.
Indeed in this model
the singularity in Fourier space lies on the surface
$\ |p|\ = \ \sqrt E$, just as for the interacting Fermions
of the previous subsection. The perturbative expansion
for the averaged Green's function is a resolvent 
expansion in which the integration over the potential
creates structures similar to the $\phi^{4}_{4}$ graphs, 
as for self avoiding polymers; the main difference is in the combinatorics
for the graphs, which is the one of a ``$N=0$ component'' theory
(in particular vacuum parts are forbidden).

The main mathematical question of physical interest 
is to prove the existence or non-existence of 
localized states. This in turn
amounts to study the behavior of the average of the modulus 
square of the two-point function 

$$
lim_{\epsilon\to 0}\ \int\ |{\cal S}(x,y ; E ; \lambda V ; \epsilon)|^{2}\ 
d\mu(V)\eqno({\rm 4.6})
$$

The rigorous results in more than one dimension
are restricted up to now to the
strong coupling regime, where it has been proved that this average 
of the modulus 
square of the two-point function  decays 
exponentially [FS],[AM]. This in turn implies localization 
of all states. 
 
{\it  At small coupling}  it is expected,  that

-i)\ in two dimensions the average of the modulus 
square of the two-point function decays
exponentially with a rate dependent on the coupling

- ii)\ in three dimensions
it decays only like a power which means that some states are delocalized.

One expects also that for all (non zero) couplings the mean value of the
two-point function
$$
lim_{\epsilon\to 0}\ \int\ {\cal S}(x,y ; E ; \lambda V ; \epsilon)\ d\mu(V)
\eqno({\rm 4.7})$$
decreases exponentially, and that it is real analytic in $\ E$.

As a first step in this direction, we have proved [MPR]:

\medskip
\noindent{\bf Theorem 6}
{\it \  In two dimensions, there exists some $\kappa>\ 0\ \ $ 
such that for $\lambda$  small enough
$$
\int\ {\cal S}(x,y ; E ; \lambda V ; \ \lambda^{2+\kappa})\ d\mu(V)\ 
\sim\ e^{- cst(\kappa)\ \lambda^{2}\ |x-y| }
\eqno({\rm 4.8})$$
which is the expected rate of decrease .}

\medskip

To suppress the regularization $\epsilon=\lambda^{2+\kappa}$ 
in our theorem, hence to prove the exponential decay
of (4.7) is well under way.

We can only give a flavor of the arguments used in Theorem 6.
A key new ingredient is to use the rhombus rule
in two dimensions for the part of the theory in which
the momenta are close to the singularity $p^{2} =E$, to create
loops with approximate momentum conservation.
These loops are then smaller than expected because of a Ward identity
which is somewhat involved technically [MPR].

But a more immediately accessible and instructive point is to apply the
decomposition of the critical surface into discrete
angular sectors also to this problem.  
One finds that the random potential sandwiched between sector cutoffs
of a given slice
$$\ {\eta}_{\sigma}(p)\ \tilde{V}(p-q)\ {\eta}_{\sigma '}(q)
\ \simeq\ {\cal{V}}_{\sigma,\sigma'}\ \ 
{\rm for}\ \ \   p \in \sigma,\ \  q \in \sigma'
\eqno({\rm 4.9})
$$
is a random matrix whose elements are labeled by the shell
cells. In two dimensions, since all momenta in a white noise distribution
are equiprobable, and since the momentum $p-q$ of $\tilde V$
approximately determines $p$ and $q$ on the circle $\ |p|\ = \ \sqrt E$
(because of the rhombus rule), this random matrix is
approximately a self adjoint matrix with  random  {\it independently
distributed} entries
$$
{\cal{V}  }(\sigma,\sigma')\ =\ \bar{\cal{V}}(\sigma',\sigma)\ 
=\ \bar{\cal{V}}(-\sigma,-\sigma') \ .
\eqno({\rm 4.10})
$$ 
Therefore we can use the mathematical theory of such matrices
for instance to bound the deviation or probability
that eigenvalues become anomalously large [M]. 
In  three dimensions this $N$ by $N$ matrix is more complicated because
$$ V_{\sigma,\sigma'}\ =\ V_{\tau,\tau'}\ \ {\rm iff} 
\ \ \sigma+\sigma'\ =\ \tau +\tau'
\eqno({\rm 4.11})
$$ 
so that approximately
$\ \sqrt{N} $ matrix elements 
corresponding to the same  momentum transfer are equal.
The study of such matrices with nonindependent entries,
and more generally the study of the weakly coupled 3 dimensional
Anderson model remains a mathematical challenge. 

\section{Further topics}

In this last section we give a brief list of open problems for
constructive methods and rigorous renormalization group studies.

- {\bf Mass generation in the Gross-Neveu model without cutoff}. Prove
that at large number $N$ of colors, the model {\it without ultraviolet cutoff}
has a spontaneously generated mass, for instance by ``gluing together'' 
the analysis of the ultraviolet limit in [DR] and the mass generation
with ultraviolet cutoff in [KMR]. This could be called ``constructive
dimensional transmutation'', after Coleman.

- {\bf Ultraviolet limit of the nonlinear $\sigma$ model in two dimension}.
This is a long standing problem. Although there are partial results,
for instance for the hierarchical model, there is no clear rigorous
construction using asymptotic freedom of this model.
Once it is built, there is also the problem of gluing the ultraviolet
analysis to mass generation (at large $N$)
to obtain ``dimensional transmutation'' also in this model. This would
be interesting in view of the controversial issues 
raised by Patrascioiu and Seiler on this kind of model.

- {\bf Fermi liquid and BCS theory in dimension 2:} Since [FMRT1],
the full construction
of the BCS vacuum at zero temperature can be considered a sound 
mathematical constructive program,
although its technical realization is very hard, and implies a tricky
constructive analysis of the infrared problems associated with
the Golstone boson of the BCS continuous symmetry breaking.

A more accessible task is to precise the mathematical status of Fermi liquid
theory itself. This is important also in view of the debate around the
nature of high-temperature superconductivity. Fermi liquid theory
is not valid at zero temperature because of the BCS instability. Even
when the dominant electron interaction is repulsive, the Kohn-Luttinger 
instabilities prevent the Fermi liquid theory to be generically valid
until zero temperature. There are nevertheless two proposals for a 
mathematically rigorous Fermi liquid theory:

- one can block the BCS and Kohn-Luttinger 
instabilities by considering models in which the Fermi surface
is not invariant under $p \to -p$ (we suggest to call this,
for obvious reasons, the ``egg model''). In two dimensions it should be 
possible to prove (nonperturbatively) that in this case the Fermi liquid theory
remains valid at zero temperature, and the corresponding program is well under
way [FKLT]. 
This program requires to control rigorously the stability of
a non-spherical Fermi surface under the renormalization group flow,
a difficult technical issue [FST].

- a simpler proposal, advocated in [S], is to study
the Fermi liquid theory at finite temperature
above the BCS transition temperature. As seen
above, the temperature acts as an infrared cutoff on the propagator
in the field theory description of the model. Hence in this point of view,
the nontrivial theorem consists in showing
that stability (i.e. summability of perturbation theory) holds
for all temperatures higher than a certain critical temperature
whose dependence in terms of the initial 
interaction is proved to be of the correct BCS form [S].

- {\bf Fermi liquid and BCS theory in dimension 3:} Here the constructive
analog of [FMRT1] must be found first. [MR] is only a first step
in this direction. Having worked for several years with J. Magnen
on this problem, it seems to me the most beautiful I met in
constructive theory until now. It is well posed and simple. For instance
one could ask: is the sum of all
convergent contributions to the theory analytic in a finite disk? Perturbative
power counting suggests that the answer should be ``yes'', but because
of the torsion angle, the constructive problem is surprisingly hard.
Again an understanding of this constructive issue
should clear up the way for many subprograms, such as
the Fermi liquid theory in 3 dimensions for ``egg models''
or above the BCS transition, and the BCS theory itself.

{\bf Anderson Model, Anderson-Mott phase transition in dimension 3}
See the previous subsection. 

- {\bf Continuation to Minkowski space, constructive scattering theory}
Here again we feel that the key for instance to a constructive
analysis of asymptotic completeness in field theory might
be related to renormalization group analysis
around a surface, this time the mass shell.
The ``constructive return'' to Minkowski space is of course 
full of interest and of important
issues for the physics of the models (scattering, bound states, time
dependent problems). Constructive study of time dependent problems
in statistical mechanics or condensed matter theory 
is also clearly a potential field for many developments.

- In the future one could focus more systematically
on clarifying the mathematical status of the 
fast developed theories of the recent decades
(Conformal theories, Non Abelian Gauge Theories, in particular 
Supersymmetric ones, Topological Theories). Some constructive tools,
of course presumably with unexpected additions, might be of help. In
this line one can also dream of
Constructive Duality and Constructive String or Membrane Theory.

\medskip
\noindent{\bf Acknowledgments: }
I thank the organizers of this session of the Swieca School for inviting 
me for this course in spite of economical uncertainties and financial cuts.
I thank P. da Veiga for coordinating his course with mine, in order
to allow me to lecture on constructive
renormalization group. Finally I thank all my recent collaborators,
A. Abdesselam, M. Disertori, C. Kopper, G. Poirot and J. Magnen,
who participated to the elaboration of the material contained here.


\end{document}